\documentclass{article}
\usepackage{comment}

\usepackage{PRIMEarxiv}
\usepackage{subcaption}
\usepackage[utf8]{inputenc} % allow utf-8 input
\usepackage[T1]{fontenc}    % use 8-bit T1 fonts
\usepackage{hyperref}       % hyperlinks
\usepackage{url}            % simple URL typesetting
\usepackage{booktabs}       % professional-quality tables
\usepackage{amsfonts}       % blackboard math symbols
\usepackage{nicefrac}       % compact symbols for 1/2, etc.
\usepackage{microtype}      % microtypography
\usepackage{lipsum}
\usepackage{fancyhdr}       % header
\usepackage{graphicx}       % graphics
\usepackage{graphicx}
\usepackage{wrapfig}
\usepackage{amsmath}

\usepackage[backend=biber, sorting=none, natbib=true, firstinits=true,uniquename=false, doi=false,isbn=false,url=true,eprint=false, defernumbers=false, maxbibnames=99]{biblatex}

\title{NL2GDS: LLM-aided interface for Open Source
Chip Design
} 
\author{
 Max Eland $^1$,  Jeyan Thiyagalingam$^2$, Dinesh Pamunuwa$^1$, and Roshan Weerasekera$^1$ \\
  $^1$School of Electrical Electronic and Mechanical Engineering (EEME) \\
  University of Bristol, Bristol, UK\\
  \\
  $^2$Science and Technology Facilities Council (STFC)\\ 
  Rutherford Appleton Laboratory, Didcot, UK \\
  \texttt{corresponding author: roshan.weerasekera@bristol.ac.uk} \\
}

\begin{document}
\maketitle

\begin{abstract}
The growing complexity of hardware design and the widening gap between high-level specifications and register-transfer level (RTL) implementation hinder rapid prototyping and system design. We introduce NL2GDS (Natural Language to Layout), a novel framework that leverages large language models (LLMs) to translate natural language hardware descriptions into synthesizable RTL and complete GDSII layouts via the open-source OpenLane ASIC flow. NL2GDS employs a modular pipeline that captures informal design intent, generates HDL using multiple LLM engines and verifies them, and orchestrates automated synthesis and layout. Evaluations on ISCAS’85 and ISCAS’89 benchmark designs demonstrate up to 36\% area reduction, 35\% delay reduction, and 70\% power savings compared to baseline designs, highlighting its potential to democratize ASIC design and accelerate hardware innovation.

\end{abstract}

% keywords can be removed

%%%%%%%%%%%%%%%%%%%%%%%%%%%%

\section{Introduction}
Recent progress in natural language processing (NLP) and large language models (LLMs) has shown that user intent, expressed in plain language, can be translated into executable software artifacts \cite{chen_evaluating_2021,austin_program_2021}. In software engineering, natural language programming has demonstrated the potential to reduce entry barriers, increase productivity, and enable domain-specific automation. Extending these capabilities to electronic design automation (EDA) and application-specific integrated circuit (ASIC) design could fundamentally reshape hardware development workflows by providing an intuitive interface to complex backend processes.

ASIC design is inherently a multi-stage workflow, spanning from high-level hardware specification to a fabrication-ready physical layout. Typical stages include register-transfer level (RTL) design, logic synthesis, floorplanning, placement, routing, timing closure, and layout verification. Traditionally, these stages rely on proprietary EDA toolchains that are often prohibitively expensive, limiting access for many small and medium-sized enterprises (SMEs) and educational institutions. Open-source EDA frameworks such as OpenLANE \cite{shalan_building_2020,noauthor_notitle_nodate} have emerged as viable alternatives, offering end-to-end ASIC implementation flows. OpenLANE offers automated synthesis, floorplanning, placement, clock tree synthesis, and routing, and can significantly reduce the cost and resource requirements of silicon prototyping.

\begin{figure}[h!]
    \centering
    \includegraphics[width=\linewidth]{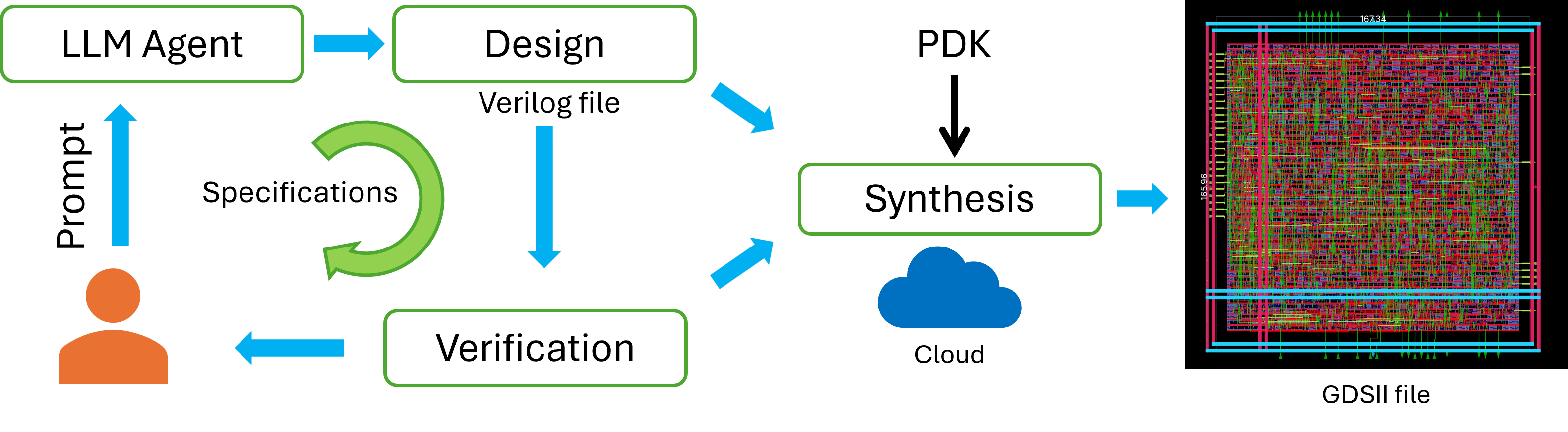}
    \caption{NL2GDS Flow}
    \label{fig:NL2GDSFlow}
\end{figure}

In this work, we introduce \textbf{NL2GDS}, a framework that generates the chip layout for digital designs directly from natural language specifications. NL2GDS combines a large language model (LLM)-based front end with the OpenLANE physical design flow \cite{shalan_building_2020,noauthor_notitle_nodate}. Users can specify hardware requirements in plain language, which the system automatically translates into synthesizable HDL, configures for OpenLANE, and produces complete GDSII layouts. By abstracting low-level implementation details, NL2GDS accelerates the progression from concept to layout, enables rapid prototyping, and supports educational applications. The overall workflow is depicted in Figure \ref{fig:NL2GDSFlow}.

Our work marks an important advance toward democratizing ASIC design, enabling agile hardware innovation, and broadening access to silicon prototyping
\begin{enumerate}
\item \textbf{Natural Language to ASIC Flow:} Introduce NL2GDS, the first framework to translate natural language hardware specifications into synthesizable RTL and full GDSII layouts using the open-source OpenLANE flow.
\item \textbf{Modular LLM-Driven Pipeline:} Employ a multi-agent architecture with retrieval-augmented generation (RAG) \cite{lewis_retrieval-augmented_2020} and chain-of-thought reasoning to automatically generate HDL, correct errors, and optimize design parameters.
\item \textbf{End-to-End Automation:} Integrate HDL generation, verification, synthesis, and layout into a fully automated cloud-based pipeline, reducing turnaround time and lowering barriers to ASIC design.
\item \textbf{Performance Gains:} Demonstrate up to 36\% area reduction, 35\% delay reduction, and 70\% power savings on ISCAS’85 \cite{Brglez1985} and ISCAS’89 \cite{brglez_combinational_1989} benchmarks, showing competitive results relative to gate-level optimized designs.
\end{enumerate}

The remainder of this paper is organized as follows. Section 2 reviews related work. Section 3
presents the system architecture, and Section 4 %IV 
describes the implementation. Section 5
 reports case studies, and Section 6
concludes with future work.

\section{Related Work and Motivation}
The rapid progress of LLMs has stimulated significant research interest in applying natural-language-driven automation to hardware design \cite{xu_llm-aided_2024}. Prior work spans (1) LLM-based HDL generation and EDA-tool-aware iterative refinement, and (2) end-to-end EDA flow orchestration. While these efforts demonstrate the feasibility of using LLMs as design copilots, none directly address the challenge of generating manufacturable GDSII layouts directly from natural language descriptions using open-source tools. We review these areas below and identify the gap that motivates our work.

\subsection{LLM-Based HDL Generation}
A conversational interface for hardware design is proposed in Chip-Chat \cite{blocklove_chip-chat_2023} demonstrating how natural-language interactions with LLMs can guide the synthesis and modification of HDL code. The system emphasizes human-in-the-loop design, enabling iterative refinement of designs through dialog and highlighting the potential of LLMs to act as design copilots. Chip-Chat identifies key challenges in mapping natural-language intent to hardware constructs as maintaining context across design conversations and integrating user feedback into the design flow. Although primarily implemented to provide a proof of concept, the framework lays the groundwork for interactive, LLM-driven hardware design and illustrates the feasibility of using conversational systems to increase accessibility and accelerate design iteration. The work proposed in \cite{xu_llm-aided_2024} presents an LLM-aided hardware design workflow that includes RTL generation, testbench synthesis, and debugging support. Their work shows that LLMs can automate many front-end design tasks, though RTL remains the upper bound of automation. \citeauthor{nakkab_rome_2024} in \cite{nakkab_rome_2024} introduce a hierarchical prompting strategy in which an LLM decomposes a hardware specification into submodules and synthesizes each component individually. This improves design modularity and correctness, enabling high-quality RTL generation from natural language input. Despite these advances, these methods focus solely on front-end design. None integrate downstream physical design or GDS production.

\subsection{End-to-End EDA Flow Orchestration}

Recent work has begun to connect LLMs to full RTL-to-GDS flows. Most notably, MCP4EDA \cite{wang_mcp4eda_2025} introduces a Model Context Protocol (MCP) interface that allows an LLM to invoke open-source EDA tools (Yosys, OpenLane, KLayout, etc.) through structured tool calls. MCP4EDA demonstrates the potential of conversational design automation: users can request synthesis, simulation, and layout operations via natural language, and the LLM can respond with synthesis scripts, layout inspection, or metric queries. This marks the first reported attempt to give LLMs operational control over a complete flow. Nevertheless, MCP4EDA remains a tool-orchestration system rather than a true natural-language-to-layout synthesis engine. It assumes the existence of an RTL implementation and relies on user intent to guide flow commands. It does not generate complete designs from high-level natural language nor perform layout-driven optimization of the desired circuit.  

\subsection{Motivation and Our Contribution}
Despite rapid progress, the literature reveals two major gaps: \textbf{Lack of backend integration in generative flows}, and \textbf{absence of direct NL to GDS synthesis}. Existing systems generate RTL but do not translate natural language into manufacturable layouts or incorporate layout-level feedback (timing, routing, DRC) into the generative loop. No prior work demonstrates the ability to take a natural-language behavioral description and automatically produce a full GDSII layout through an end-to-end LLM-driven flow. These gaps motivate the need for an approach that couples (i) high-level natural language reasoning, (ii) HDL synthesis, (iii) backend physical design via OpenLane, and (iv) tool-aware iterative refinement - all within a single automated pipeline.

\begin{figure}[h!]
    \centering
    \includegraphics[width=0.5\linewidth]{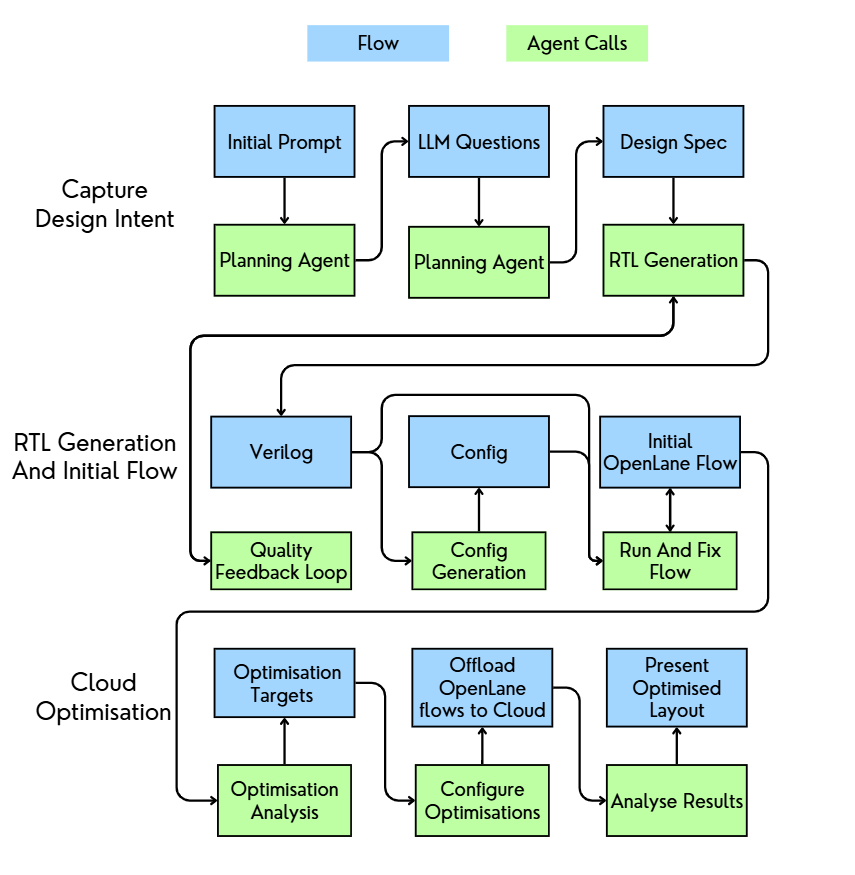}
    \caption{NL2GDS Block Diagram}
    \label{fig:Flow_diagram}
\end{figure}

In contrast to prior work, our system provides the first end-to-end natural-language to GDSII design pipeline grounded in the OpenLane flow. Specifically, our contributions are:
\begin{itemize}
\item \textbf{Full generative synthesis}: We generate RTL, constraints, and physical design configurations directly from natural language prompts.
\item \textbf{Backend-aware refinement}: The LLM ingests OpenLane metrics and refines both RTL and flow configuration iteratively.
\item \textbf{Unified conversational interface}: Unlike MCP-based orchestration \cite{wang_mcp4eda_2025} our system integrates generation, tool invocation, and physical optimization in a single natural-language loop.
\item \textbf{Democratized chip design}: Users with limited HDL or EDA expertise can produce manufacturable layouts through high-level descriptions alone.
\end{itemize}
In doing so, our system bridges the long-standing disconnect between natural-language specification and manufacturable chip layout, expanding the frontier of LLM-driven design automation and addressing key limitations of prior NL-to-RTL and tool-orchestration-only frameworks.

%%%%%%%%%%%%%%%%%%%%%%%%%%%%%%%
\section{System Overview}
The NL2GDS architecture (see Figure \ref{fig:Flow_diagram}) is a fully modular, AI-driven platform that automates the entire RTL-to-GDSII hardware design flow. We utilise LLM-driven agents for orchestration and have developed specific tool calls that enable each agent to interact with the OpenLane flow and cloud processing resources. The system consists of four core, tightly integrated components, each addressing specific barriers of traditional script-based EDA.

\subsection{Web Based Front End}
\begin{figure*}[h!]
    \centering
    \includegraphics[width=0.9\linewidth]{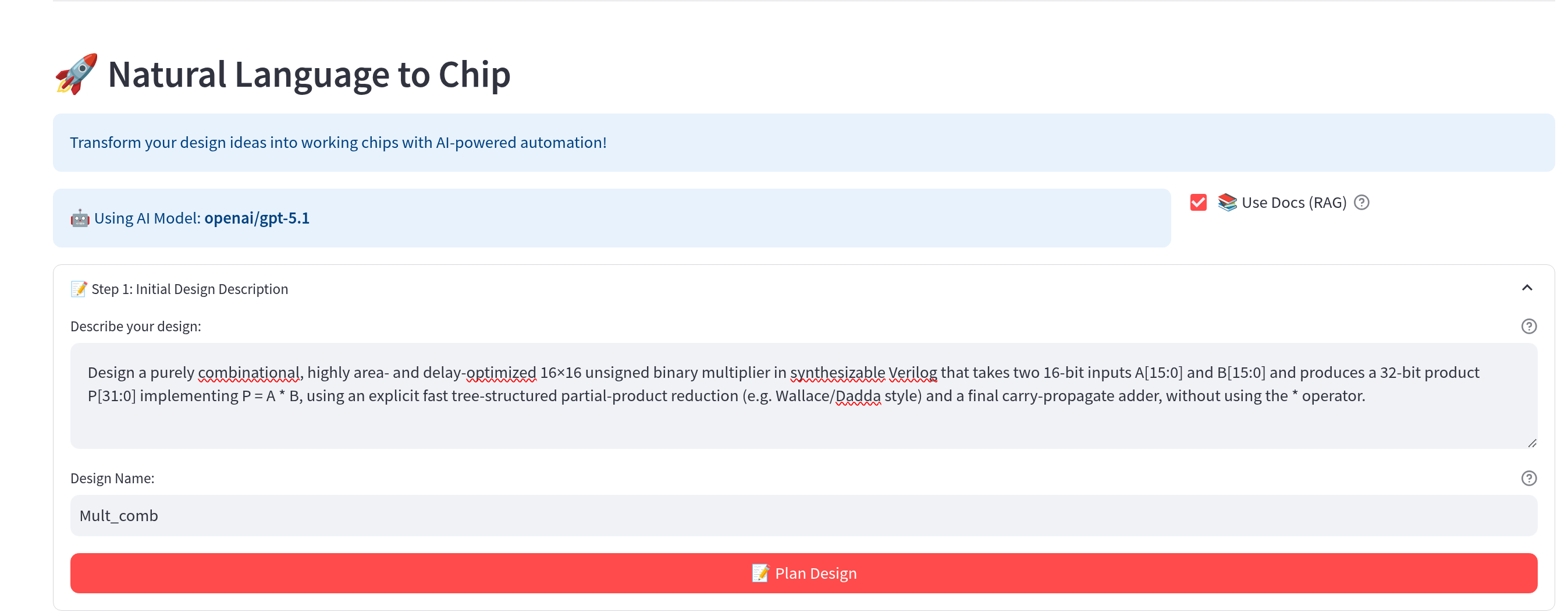}
    \caption{NL2GDS Front End Initial Prompt}
    \label{fig:Front_end}
\end{figure*}

Figure \ref{fig:Front_end} shows the first dedicated browser interface for OpenLane, offering seamless integration with internal GUIs from KLayout and OpenRoad. Developed using the Streamlit Python library, the interface eliminates CLI/scripting complexities, enabling faster design iteration and optimisation. We lower the knowledge barrier to entry whilst still producing high quality, foundry ready designs. The platform is fully dockerized and runs in the cloud, enabling users to access the tool from any location, regardless of their operating system or hardware.

\subsection{Agentic Approach}
The NL2GDS system utilizes a team of specialized agents to automate each stage of the design flow, producing higher quality outputs compared to single shot LLM calls. They interface with OpenLane, gathering context from synthesis, floor-planning, timing, and PPA (Power, Performance, and Area) logs. With thousands of files generated from each OpenLane run, LLMs are vital in the effective synthesis of this knowledge. The Agents are specifically optimized to understand user intent,  generate HDL, fix errors specific to the OpenLane flow, and provide post layout optimization suggestions. They have tool calls such as RAG context checks, file write/read privileges, and interfacing with distinct OpenLane flow sections.   

\subsection{Retrieval Augmented Generation}
RAG allows LLMs to access relevant information outside of their fixed context window. The NL2GDS system employs RAG to access project specific documentation and successful designs in order to enhance configuration file generation inside of OpenLane. Base LLMs do not have a good understanding of the approximately 800 design parameters that control the OpenLane flow. By giving them targeted access to these we dramatically improve reliability and creativity when tackling new designs. This also allows for targeting error fixing when dealing with OpenLane error codes and intelligent regeneration to fix these. 

\subsection{Cloud Backend}
The flow is fully automatic, going from natural language to foundry ready layout without designer intervention. It prioritises user time, automatically fixing errors that arise during the flow. The system leverages cloud compute to enable rapid design exploration, covering parameter sweeps and layout optimisations. Designers define PPA objectives and the system orchestrates parallel runs to iteratively optimize the design. This democratizes silicon prototyping at a fraction of the cost for users lacking on-premise compute resources.

%%%%%%%%%%%%%%%%%%%%%%%%%%%%%%%
\section{Implementation}
This implementation addresses three primary bottlenecks commonly associated with agentic EDA flows: Verilog generation, missing flow context, and compute resource limitations that hinder optimization progress.

\subsection{Chain of Thought}
The chain-of-thought (CoT) process is fundamental for agentic applications. Rather than adhering to an initial response, tasks are decomposed into smaller, more manageable components, enabling the completion of a sequence of subtasks that collectively achieve a broader objective. This methodology transforms linear workflows into adaptive processes and provides greater control over tool invocation by dividing the sequence into precise tasks. The CoT process is implemented throughout the system for planning, Verilog generation, verification, automated error correction, and optimization. CoT approaches have been shown to significantly enhance performance in AI systems. For instance, CoT prompting has shown substantial increases in code quality, such as a 63\% relative improvement in Pass@1 accuracy on the HumanEval benchmark \cite{yang2024chainofthoughtneuralcodegeneration}.

An optimized planning agent is utilized to capture user intent. The process begins with a high-level design prompt, followed by LLM-generated targeted questions for the user regarding functionality, inputs and outputs, architecture, and PPA optimization priorities. Rather than taking over the design process, the LLM collaborates with the user until a comprehensive understanding is achieved. The LLM then produces a design specification for the Verilog Generation Agent tailored to its capabilities. This methodology has demonstrated high responsiveness to precise specifications. For Verilog generation, a feedback loop is implemented that incorporates additional logic checks by a separate agent and automated invocation of Verilator in lint mode. This process identifies coding errors, poor practices, compatibility issues, and other errors that could prevent the OpenLane flow from executing. Only after the code is thoroughly checked and cleared is it presented to the user for verification prior to synthesis. This implementation was validated on the VerilogEval dataset using a subset of 10 challenging test cases \cite{pinckney2024revisitingverilogevalnewerllms}. This subset focuses on problems that require generating intricate FSMs (like fsm\_hdlc, conwaylife, and fsm\_ps2data), handling wide vector operations (popcount255), and implementing specific hardware algorithms (like the gshare branch predictor and lfsr32).

The CoT implementation was initially evaluated on this subset using Gemini 2.5 pro, both with and without the NL2GDS architecture. The integration of Gemini 2.5 pro and NL2GDS resulted in an 11\% increase in successful testbench checks across the subset compared to the base model. Subsequently, prompt 256 from the VerilogEval dataset, identified as one of the most challenging FSM tasks with 200,000 testbench checks, was employed to demonstrate universality across leading LLMs at the time of writing. Table \ref{tab:tb-pass-rates} presents the base success rate as the percentage of successful testbench checks, compared to the performance achieved by applying the NL2GDS architecture to each LLM. The results indicate that NL2GDS consistently improves Verilog generation for all models. 

\begin{table}[htbp]
\centering
\caption{Testbench pass rates from generated Verilog with and without NL2GDS COT.}
\label{tab:tb-pass-rates}
\begin{tabular}{lrr}
\hline
Model & Baseline pass rate (\%) & With NL2GDS CoT (\%) \\
\hline
Gemini 2.5 Pro & 3.79 & 83.70 \\
GPT-5 & 17.52 & 100.00 \\
Sonnet 4.5 & 99.88 & 100.00 \\
Grok-4 Fast & 0.00 & 18.01 \\
MiniMax M2 & 0.00 & 95.62 \\
\hline
\end{tabular}
\end{table}

The results indicate substantial improvements in code generation, particularly in adhering to precise instructions regarding functionality and input/output naming to satisfy unseen test benches from a natural language prompt. Additionally, the model generated syntactically accurate code even when base models failed, demonstrating robust self-correction.

\subsection{RAG in the NL2GDS Flow}

LLMs do not have a good understanding of the OpenLane EDA flow. They struggle to generate synthesisable configuration files required for the flow to start due to a lack of syntax knowledge and how each parameter affects the design. To overcome this, we have used a design specific RAG system to inject relevant OpenLane documentation, parameter guides and successful examples into the prompts. 

For the initial run, we parse metrics from the generated Verilog and design specification into an initial algorithm that detects complexity and key design specification parameters to achieve PPA. This queries successful designs to generate a domain specific initial config. The flow is orchestrated and fixed by the LLM until completion. From here we have the required logs to begin optimisation. 

We define the extraction of critical design issues via a detection operator:

\begin{equation}
    \mathcal{I} = f_{\mathrm{detect}} (\mathcal{M}, \mathcal{E}) = \left\{ i_1, i_2, \ldots, i_n \right\}
\end{equation}

Where $\mathcal{I}$ encodes detected timing, area, routing, and DRC violations from extracted metrics($\mathcal{M}$) and Errors($\mathcal{E}$).

The implementation of $f_{\mathrm{detect}}$ includes heuristic rules that identify negative setup or hold slack as timing violations, high placement utilization as potential area congestion, and nonzero DRC/LVS counts as physical or connectivity errors. Each detected issue is encoded with its category, severity, and location within the flow. These structured issues ($\mathcal{I}$), along with the current configuration ($\mathcal{C}$) and, if specified, a user-defined optimization goal ($\mathcal{G}$), are used to generate targeted queries:

\begin{equation}
    \mathcal{Q} = f_{\mathrm{query}} (\mathcal{I}, \mathcal{C}, G) = \left\{ q_1, q_2, \ldots, q_k \right\}
\end{equation}

Queries are formulated to retrieve documentation and validated solutions directly relevant to specific issues, such as “OpenLane timing optimization CLOCK\_PERIOD violation” or “Area reduction via FP\_CORE\_UTIL”. For each query, the system presents the top-$\mathcal{N}$ most frequently referenced parameter explanations, documentation segments, and configuration values.

\begin{equation}
    \mathcal{D}_{q_j} = \operatorname{Retrieve}(q_j, \mathcal{N}), \quad
    \mathcal{D}_{\mathrm{RAG}} = \bigcup_{j=1}^{k} \mathcal{D}_{q_j}
\end{equation}

The combined set of retrieved documents $\mathcal{D}_{\mathrm{RAG}}$, together with run metrics, logs, configuration state, and optimization history, is incorporated into the prompt to enable the LLM to generate syntactically valid and design-aware configuration changes.

\begin{align}
\mathcal{O}_{RAG} &= \operatorname{Concat}\big(\mathcal{M},\,\mathcal{E},\,\mathcal{H},\,\mathcal{C},\,\mathcal{D}_{\mathrm{RAG}},\,\mathcal{G}\big)
\end{align}

\subsection{Parallelism}
OpenLane is traditionally designed to be run as independent Operating System processes via the command line instead. NL2GDS relies substantially on python for orchestrating the flow; however, relying on Python's Global Interpreter Lock (GIL) can restrict true parallelism for CPU-bound workloads \cite{10596487}. Since OpenLane is itself an external binary, we decoupled it from Python’s runtime limitations. We use an I/O parallel approach for orchestration due to the large number of read/write functions. This only needs to handle the data transfer instead of CPU intensive EDA tasks. We then launch each OpenLane run into a subprocess to run simultaneously on multicore servers or cloud nodes, each consuming substantial resources and progressing independently, unconstrained by Python’s threading model.

A total of 100 flows were executed on the OpenLane’s unsigned serial/parallel multiplier (SPM) design. In sequential mode, each run required 132 seconds to complete the flow. The parallel implementation achieved an average runtime of 37.5 seconds per flow, representing a 3.46-fold increase in speed compared to sequential execution. Furthermore, all design configurations were automatically generated and launched by the optimization agent, which analyzed targeted logs to identify potential improvements. When accounting for the manual time typically required for designers to analyze logs and modify configurations, the overall productivity gain is even more substantial.

\section{Evaluation}
This section presents the PPA optimization capabilities of the NL2GDS flow, as well as the complexity of designs it can generate. The evaluation uses the Open Source Skywater 130nm PDK. A selection of ISCAS Verilog benchmark designs, including both combinational and sequential circuits, serves as the basis for assessment. We incorporated a combination of established ISCAS benchmark circuits and designs generated through the NL2GDS flow as inputs to the OpenLane physical design toolchain.  This allowed us to systematically characterize each design in terms of fundamental implementation metrics such as silicon area, critical-path delay, and estimated power consumption. By grounding our evaluation in these quantifiable metrics, we developed a robust framework that directly aligns with the broader goals of our work. The evaluation framework focuses on measurable outcomes that align with our primary objectives: competing with proprietary benchmarks from purely natural language prompting, reducing time to layout, and achieving optimized designs at significantly lower cost.

\subsection{ISCAS vs NL2GDS}
The quality of designs generated by NL2GDS was evaluated by providing instructions to create functionally equivalent ISCAS circuits. Each NL2GDS design was explicitly told to preserve identical input and output specifications and functionality, as verified by a dedicated test bench. The ISCAS circuits were processed through the OpenLane flow, after which NL2GDS was used to generate the functionally equivalent designs. To evaluate the performance, we have used a ratio of  the NL2GDS performance vs ISCAS as given by $\mathcal{R}$. Ratios lower than 1 show NL2GDS beating the ISCAS implementation for the relevant metric.
%traditional benchmarks.   

\begin{equation}
\mathcal{R} = \frac{\text{NL2GDS}}{\text{ISCAS}}
\end{equation}

\begin{figure}[h]
    \centering
    \includegraphics[width=0.64\linewidth]{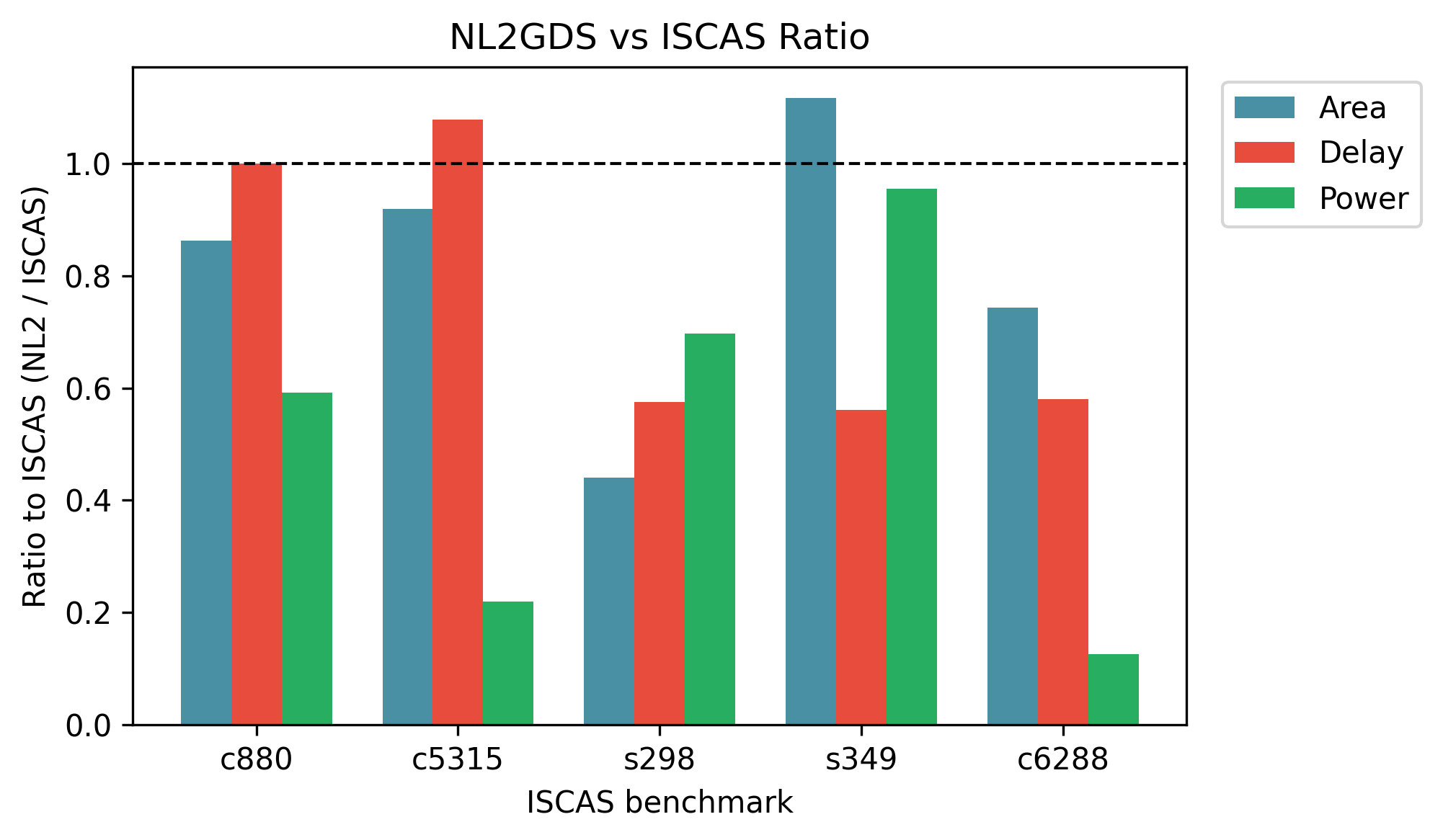}
    \caption{Comparison of NL2GDS Vs ISCAS for Area, Delay and Power Metrics}
    \label{fig:Benchmarks}
\end{figure}

\begin{figure}[h]
     \centering
     \begin{subfigure}[b]{0.48\linewidth}
         \centering
         \includegraphics[width=\linewidth]{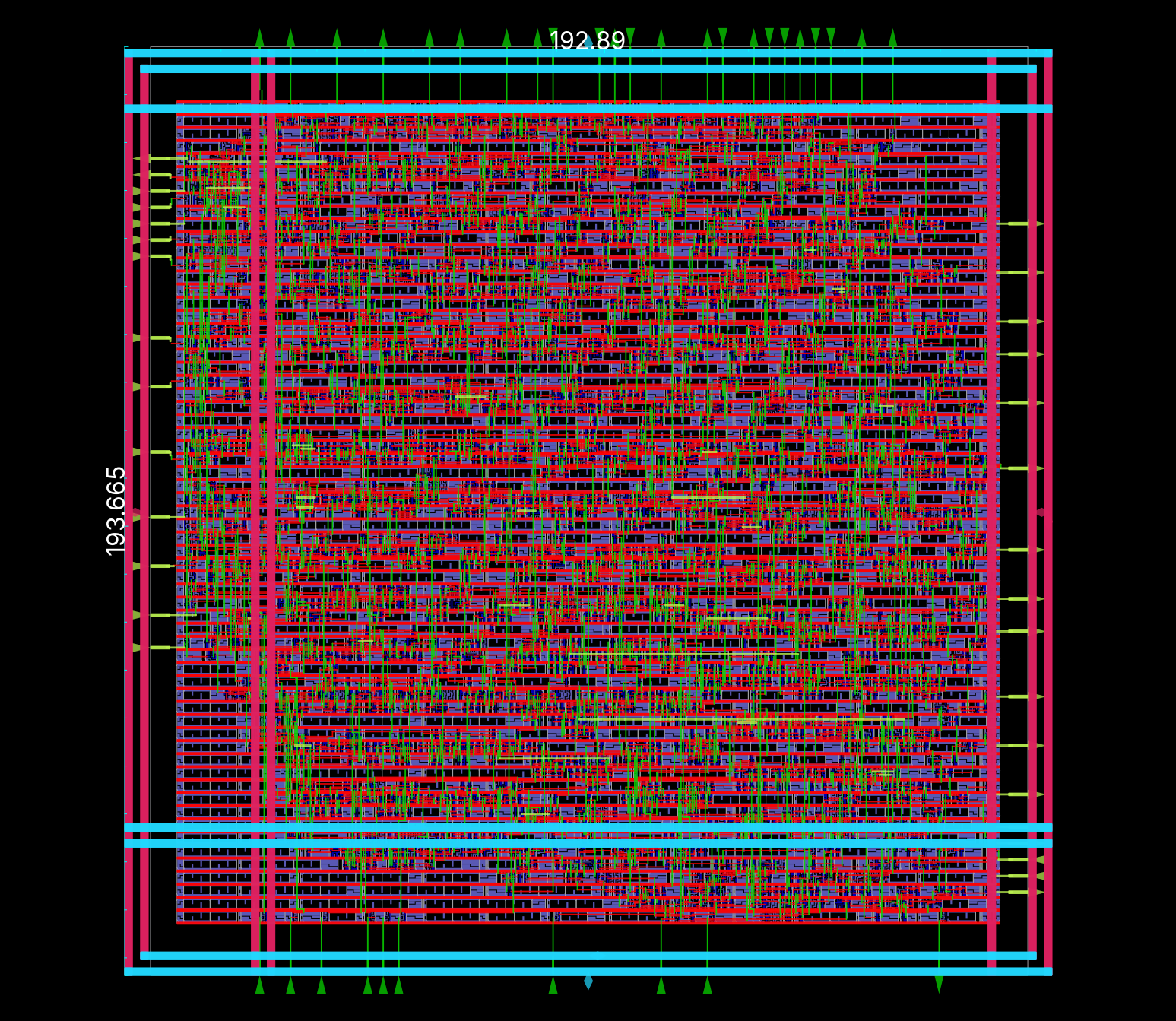}
         \caption{ISCAS input}
         \label{fig:Iscas Mult}
     \end{subfigure}
     \hfill
     \begin{subfigure}[b]{0.48\linewidth}
         \centering
         \includegraphics[width=\linewidth]{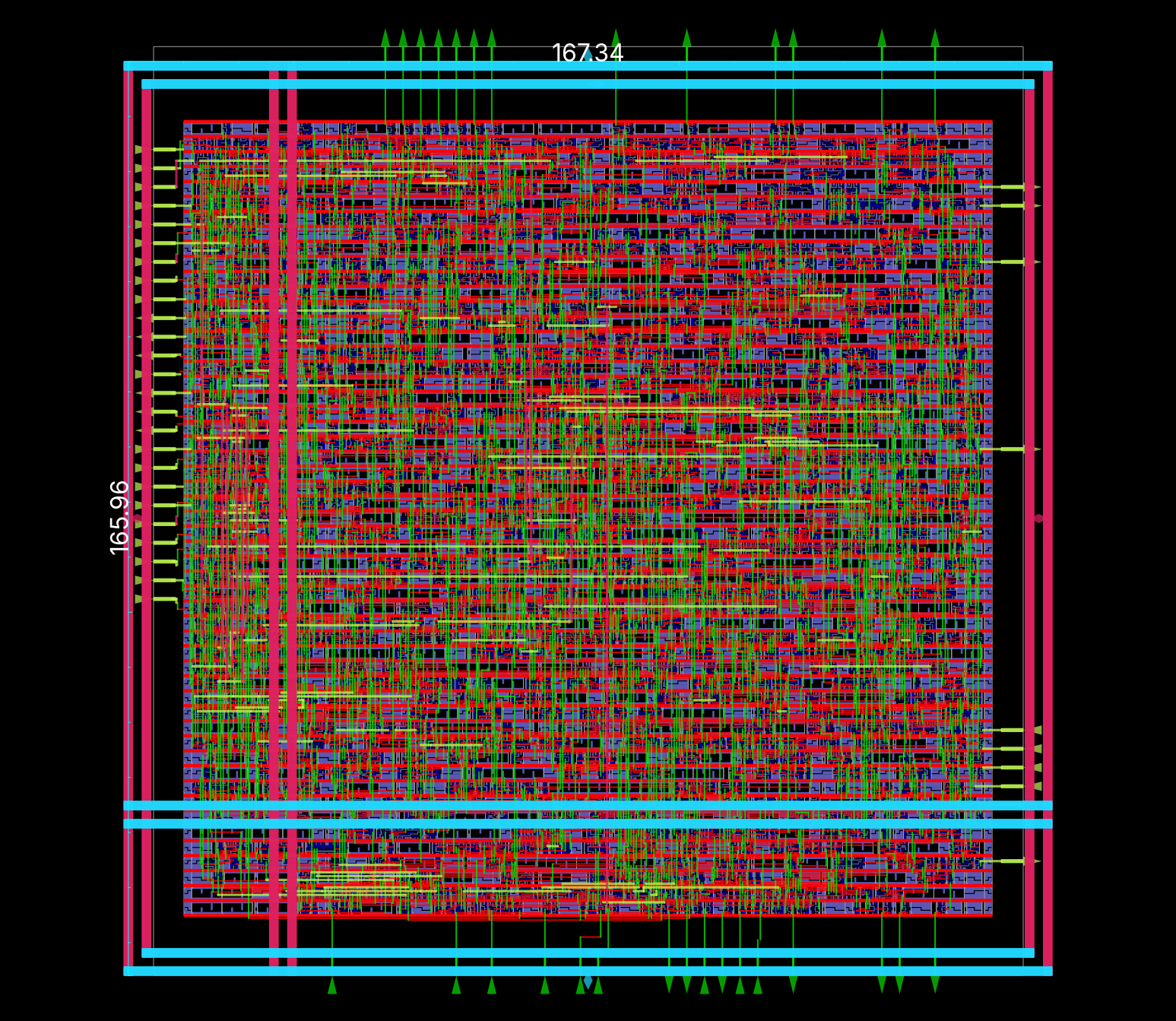}
         \caption{NL2GDS-generated Verilog}
         \label{fig:NL2GDS Mult}
     \end{subfigure}
\caption{Layouts generated for a 16×16 combinational multiplier in OpenROAD using (a) ISCAS input (chip size: 192.89um × 193.665um) and (b) NL2GDS-generated Verilog code (167.34um × 165.96um).}
\end{figure}

The NL2GDS system not only generated the designs, but also identified optimized layouts for each through self-optimization. The ISCAS designs are highly optimized and written in gate-level Verilog. NL2GDS better results across every design for all PPA metrics bar two instances out of 15. As an example, figures \ref{fig:Iscas Mult} and \ref{fig:NL2GDS Mult}, show the two c6288 designs from ISCAS benchmark circuits \cite{Brglez1985}; the NL2GDS system achieves a 25.7\% reduction in die area compared to the ISCAS benchmark implementation. 

\begin{table*}[t]
\centering
\caption{Combined Results for Baseline and Optimised NL2GDS Designs}
\label{tab:PPA_results} 
\resizebox{\columnwidth}{!}{%
\begin{tabular}{|l|r|r|r||r|r|r||r|r|r|}
\hline
& \multicolumn{3}{c||}{Area (A) in µm$^2$} & \multicolumn{3}{c||}{Delay ($t_d$) in ps} & \multicolumn{3}{c|}{Power (P) in $\mu$W)} \\ \cline{2-10}
\textbf{Design} & \textbf{Base-} & \textbf{NL2GDS} & \textbf{$\Delta A$ (\%)} & \textbf{Base- } & \textbf{NL2GDS} & \textbf{$\Delta t$ (\%)}  & \textbf{Base-} & \textbf{NL2GDS} & \textbf{$\Delta P$ (\%)} \\ 
 & \textbf{line} &  &  & \textbf{line } &  &  & \textbf{line} &  &  \\ \hline

Combinatorial 8-bit ALU (C880)      & 7229  & 4650  & -35.68 & 15284  &  9955  & -34.86 & 77    & 32    & -58.44 \\
16-bit error detector/corrector (C2670) & 5833  & 3883  & -33.43 & 15046  &  9877  & -34.35 & 266   & 81    & -69.55 \\
9-bit ALU (C5315)                  & 67600 & 62500 & -7.54  & 18284  & 12955  & -29.15 & 127   & 112   & -11.81 \\
Sequential traffic-light controller (S298) & 1727  & 1564  & -9.44  & 6970   & 5737   & -17.69 & 98    & 64    & -34.69 \\
4x4 Add-Shift Multiplier (S349)    & 5863  & 4472  & -23.73 & 8141   & 5607   & -31.13 & 206   & 132   & -35.92 \\
Digital Fractional Multiplier (S838)   & 29576 & 19683 & -33.45 & 12268  & 9758   & -20.46 & 3319  & 1974  & -40.52 \\
16x16 Multiplier (C6288)           & 61317 & 27771 & -54.71 & 21607  & 17765  & -17.78 & 1505  & 1213  & -19.40 \\
16x16 Multiplier (Pipelined)       & 90979 & 58071 & -36.17 & 11720  & 7584   & -35.29 & 4521  & 2409  & -46.72 \\
\hline
\end{tabular}}
\end{table*}

\subsection{NL2GDS Self Optimisation}

From here we expanded our test case to include more complex designs. Table \ref{tab:total-runs-cost} presents the number of OpenLane runs and the total time required from the initial prompt to the completion of the optimized layout. The user specifies the full functionality to the LLM by responding to its queries, eliminating the need for a netlist. The results demonstrate that rapid optimization is achieved in a fraction of the time required by traditional development methods. Additionally, the entire workflow incurs minimal computational and LLM-related costs.

\begin{table}[h!]
    \centering
    \caption{Total NL2GDS runs and cost per design.}
    \label{tab:total-runs-cost}
    %\resizebox{\columnwidth}{!}{%
    \begin{tabular}{lccc}
        \hline
        Design &
        Runs &
        Time [min] &
        LLM + compute cost [\$] \\
        \hline
        C880 &
        75 &
        20 &
        0.48 \\
        C2670 &
        100 &
        14 &
        0.32 \\
        C5315 &
        72 &
        26 &
        0.53 \\
        S298 &
         85 &
         12 &
         0.13 \\
        S349 &
        75 &
        18 &
        0.27 \\
        S838 &
        75 &
        26 &
        0.51 \\
        16x16 Mult (C6288) &
        50 &
        28 &
        0.56\\
        16x16 Mult (Pipelined) &
        48 &
        30 &
        0.56
        \\
        \hline
    \end{tabular}%
    %}
    
\end{table}

The design metrics for the benchmark circuits are shown in Table \ref{tab:PPA_results} demonstrating the capability of rapid design exploration using the NL2GDS system. The baseline is the first successful run of the OpenLane flow and the  optimized design with PPA metrics after the number cloud runs in Table \ref{tab:total-runs-cost}.  

All designs show reduced die area, with savings ranging from 7.54\% (9-bit ALU) to 54.71\% (16×16 multiplier). Larger arithmetic circuits, especially multipliers, benefit the most, achieving reductions above 35\%, while smaller control-oriented designs show modest gains. There are significant timing improvements, with reductions between 17.69\% and 35.29\%. Power savings are substantial, ranging from 11.81\% to 69.55\%. Arithmetic-heavy designs such as multipliers and ALUs generally see reductions above 35\%, confirming the effectiveness of NL2GDS in lowering dynamic power.

\subsection{Multiplier Case study}
The two multiplier designs demonstrate the capability to rapidly prototype various architectures. One design was implemented as a purely combinatorial circuit, directly targeting the ISCAS benchmark. The system was then instructed to maintain functionality while introducing pipelining to enhance speed, accepting increased area and switching power as trade-offs. The results indicate a 10ns reduction in clock period for the sequential design, accompanied by a twofold increase in power consumption and a 2.09-fold increase in area due to the additional pipelining circuitry. These outcomes align with anticipated trade-offs. The total time required to complete these tests, including natural language input, tapeout generation, and 98 optimization runs for both designs, was only one hour. Additionally, the total cost amounted to \$1.12 for both LLMs and cloud services. These findings highlight the significant potential of the NL2GDS system to enable designers and researchers to efficiently prototype and evaluate design trade-offs with comprehensive metrics, without writing code.

\begin{figure}[htbp]
    \centering
    \includegraphics[width=0.9\linewidth]{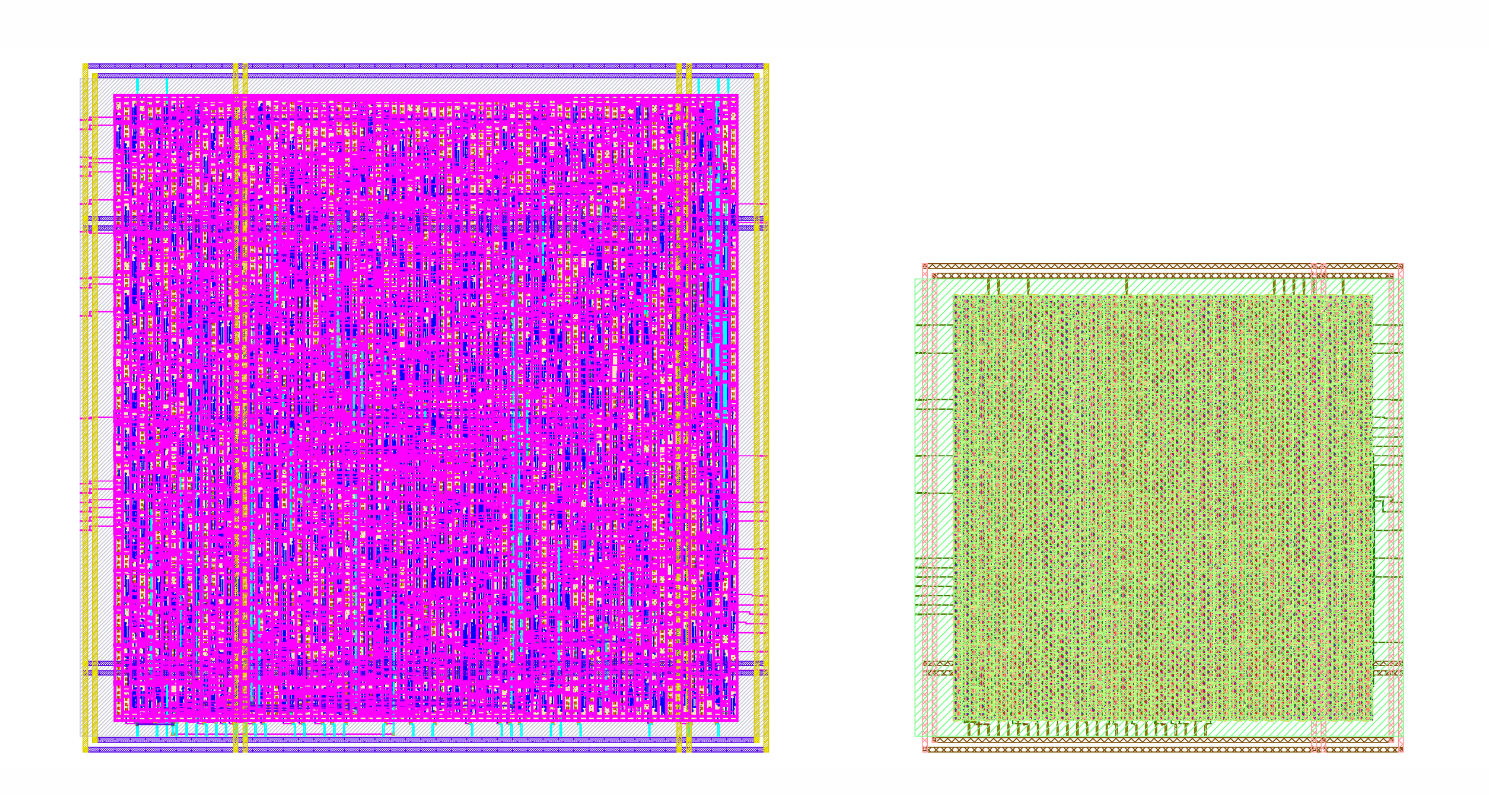}
    \caption{GDSII layouts generated for Pipelined vs Combinatorial 16x16 Multipliers in KLayout}
    \label{fig:Multiplier Comaprison}
\end{figure}

%%%%%%%%%%%%%%%%%%%%%%%%%%
\section{Conclusion}
The proposed NL2GDS flow demonstrates the ability to rapidly translate natural language specifications into GDSII layouts, bridging the gap between high-level design intent and physical implementation. It generates reliable Verilog code that enhances the performance of state-of-the-art models and introduces a novel RAG-based system for producing optimized configuration files—a capability not previously achieved by LLMs. Combined with newly developed tool calls for seamless interaction with the OpenLane flow, NL2GDS enables the creation of optimized, cloud-based designs that are competitive with gate-optimized ISCAS benchmarks.

Experimental results confirm the effectiveness of NL2GDS in delivering compact, energy-efficient, and high-performance designs. Area reductions range from 25.29\% to 54.43\%, while power savings vary between 11.81\% and 69.55\%, with arithmetic-heavy circuits such as multipliers and ALUs achieving reductions above 35\%. Critical path delays also improve significantly, with reductions of 25.29\% to 44.16\%, demonstrating that NL2GDS not only lowers dynamic power but also enhances timing performance. Overall, NL2GDS represents a transformative approach for automated hardware design, particularly suited for edge-AI and IoT applications where efficiency and rapid deployment are paramount.

%Although NL2GDS has demonstrated significant capabilities in translating natural language to GDSII, the subsequent objective is to implement complete designs by developing each module individually, such as ALUs, SRAM, and FSMs, and integrating these macros into a unified architecture like a RISC-V CPU. To achieve this, custom scripts are being developed to incorporate the Google DeepMind AlphaChip project, which has demonstrated a 6.2\% reduction in wirelength compared to human experts \cite{mirhoseini2021graph}. The project has been containerized using the open-source AlphaChip binaries, and custom scripts are being written to assist certain floorplanning steps within the OpenLane flow with reinforcement learning.

%%%%%%%%%%%%%%%%%%%%%%%%%%%%%%%%%%%%%%
\section{Acknowledgements}
We gratefully acknowledge the funding support provided through the University of Bristol Summer Internship scheme for Max Eland. 

This work was also supported by the i-EDGE project, which has received funding from the European Union (grant number 101092018), the Swiss State Secretariat for Education, Research and Innovation (SERI) and UK Research and Innovation (UKRI) under the UK government's Horizon Europe funding guarantee (grant numbers 10061130 and 10063023).

%\printbibliography

\end{document}